\documentclass[conference]{IEEEtran}

\usepackage[cmex10]{amsmath}
\usepackage{amssymb}
\usepackage{amsthm}
\theoremstyle{plain}
\newtheorem{theorem}{Theorem}
\newtheorem{proposition}[theorem]{Proposition}
\usepackage{cite}

\begin{document}

\title{Generalized Cram\'{e}r-Rao relations for non-relativistic quantum systems}

%\author{J.S. Dehesa, A.R. Plastino, P. S\'anchez-Moreno and C. Vignat}

\author{\IEEEauthorblockN{J.S. Dehesa\IEEEauthorrefmark{1}\IEEEauthorrefmark{3},
A.R. Plastino\IEEEauthorrefmark{1}\IEEEauthorrefmark{3},
P. S\'anchez-Moreno\IEEEauthorrefmark{1}\IEEEauthorrefmark{4} and
C. Vignat\IEEEauthorrefmark{2}}
\IEEEauthorblockA{\IEEEauthorrefmark{1}Instituto 'Carlos I' de F\'{i}sica Te\'{o}rica y Computacional, Universidad de Granada, Granada, Spain}
\IEEEauthorblockA{\IEEEauthorrefmark{2}Laboratoire des Signaux et Syst\`{e}mes, Supelec, France}
\IEEEauthorblockA{\IEEEauthorrefmark{3}Departamento de F\'{i}sica At\'omica, Molecular y Nuclear, Universidad de Granada, Granada, Spain}
\IEEEauthorblockA{\IEEEauthorrefmark{4}Departamento de Matem\'atica Aplicada, Universidad de Granada, Granada, Spain}}

\date{\today}

\maketitle

\begin{abstract}
The Cram\'{e}r-Rao product of the Fisher information $F[\rho]$ 
and the variance $\langle \mathbf{x}^2\rangle\equiv\int \mathbf{x}^2\rho(\mathbf{x})d\mathbf{x}$ of a probability density $\rho(\mathbf{x})$, defined
on a domain $\Delta \subset \mathbb{R}^D$, is found to have a minimum value 
reached by the density associated with the ground state of the harmonic oscillator in $\Delta$, when $\Delta$  is an unbounded domain. If $\Delta$ is bounded, the minimum value of the Fisher information is achieved 
by the ground state of the quantum box described
itself by this domain.
\end{abstract}

\section{Introduction}

The Cram\'er-Rao inequality belongs to a natural family of information-theoretic inequalities \cite{cover_thomas,johnson_04,bercher.is09} which play a relevant role in a great variety of scientific and technological fields ranging from probability theory \cite{cramer_46,renyi_70}, communication theory \cite{dembo.itit91}, signal processing \cite{kay_93} and approximation theory \cite{dehesa.jcam06} to quantum physics of $D$-dimensional systems with a finite number of particles \cite{romera.pra94,dehesa.jpa07,angulo_10}. Recently the general Cram\'er-Rao inequality \cite{cover_thomas,dembo.itit91}
\begin{equation}
F\langle \mathbf{x}^2\rangle \ge D^2,
\label{eq.cramerrao}
\end{equation}
valid for all ground and excited states of any $D-$dimensional quantum system, has been substantially improved \cite{dehesa.jpa07} in the case of centrally-symmetric potentials as
\[
F\langle \mathbf{x}^2\rangle \ge 4\left(1-\frac{2|m|}{2l+D-2}\right)\left(l+\frac{D}{2}\right)^2
\]
where the 
%removed non-negative
integer numbers $(l,m)$ denote the hyperquantum orbital and magnetic quantum numbers having the values $0 \le l \le n-1$  and $-l \le m \le  + l,$ the integer $n$ being the hyperquantum principal number.
%removed , equal to $1,2,\ldots$). 
It is worth pointing out that the lower bound to the Cram\'er-Rao product is equal to $D^2$ 
% modified here 
when $l=0$, i.e. for states $s$.

Moreover, the Cram\'er-Rao inequality has been shown to be closely connected to the Heisenberg uncertainty inequality by various authors \cite{dembo.itit91}. In particular, it is fulfilled that
\[
F\langle \mathbf{x}^2\rangle \ge 4\left(1-\frac{2|m|}{2l+D-2}\right)\langle r^2\rangle\langle p^2\rangle
\]
is valid for central potentials \cite{dehesa.jpa07}. Since the Cram\'er-Rao inequality (\ref{eq.cramerrao}) is saturated by Gaussian probability densities, it can be interpreted as a measure of the amount of non-normality of the quantum-mechanical probability density which describes the involved quantum state of the physical system under consideration \cite{dembo.itit91}.

This work, which has been motivated by the recent findings in \cite{bercher.is09}, provides the best Cram\'er-Rao lower bound for general $D$-dimensional systems on unbounded domains in Section \ref{sec2}. It is shown that this lower bound is reached for the probability density characterizing the oscillator ground state, and that its value is controlled by the corresponding ground state energy. Next, in Section \ref{sec3}, we show that the minimal value of the Fisher information of general $D$-dimensional systems defined on bounded domains  is achieved for the ground state of the quantum box described by this region.

\section{Unbounded $D$-dimensional domains}
\label{sec2}
\subsection{Notations and problem}
Let us consider an unbounded domain $\Delta$ in $\mathbb{R}^D$. The problem consists in finding the normalized probability density $\rho(\mathbf{x})=u^2(\mathbf{x})$, with $\mathbf{x}\in\Delta$, that minimizes the Fisher information with the constraints %of the given value
that the variance  $\langle \mathbf{x}^2\rangle$ has a given value, and that $u(\mathbf{x})=0$ $\forall \mathbf{x}\in \mathcal{D}(\Delta)$, the frontier of $\Delta$. The Fisher information will be denoted as $F[\rho]$ or simply $F$ when there is no ambiguity about the considered distribution; it is defined as

\[
F[\rho] = 4\int_\Delta |\boldsymbol\nabla u(\mathbf{x})|^2d\mathbf{x}.
\]

The Lagrangian of this problem is
\begin{eqnarray*}
\mathcal{L}=4\int_\Delta |\boldsymbol\nabla u(\mathbf{x})|^2d\mathbf{x}
+\alpha\left[\int_\Delta u^2(\mathbf{x})d\mathbf{x}-1\right] \\
+\beta\left[\int_\Delta \mathbf{x}^2 u^2(\mathbf{x})d\mathbf{x}-\langle\mathbf{x}^2\rangle\right],
\end{eqnarray*}
where $d\mathbf{x}=\prod_{i=1}^{D}dx_{i}$, $\mathbf{x}^2=\sum_{i=1}^{D}x_{i}^2$, and
\[
|\boldsymbol\nabla u(\mathbf{x})|^2=\sum_{i=1}^D\left(\frac{\partial u}{\partial x_i}\right)^2.
\]
The associated Euler-Lagrange equation states that
\[
\sum_{i=1}^D\frac{\partial}{\partial x_i}\frac{\partial \mathcal{L}}{\partial\left(\frac{\partial u}{\partial x_i}\right)}-\frac{\partial \mathcal{L}}{\partial u}=0,
\]
which yields the differential equation
\[
8\boldsymbol\nabla^2u(\mathbf{x})-2\alpha u(\mathbf{x})-2\beta \mathbf{x}^2u(\mathbf{x})=0.
\]
This equation, together with the boundary condition $u(\mathbf{x})=0$, $\forall \mathbf{x}\in \mathcal{D}(\Delta)$, %is equivalent to
coincides with the Schr\"odinger equation
\[
-\frac12\boldsymbol\nabla^2u(\mathbf{x})+V(\mathbf{x})u(\mathbf{x})=-\frac{\alpha}{8}u(\mathbf{x})
\]
with the potential
\begin{eqnarray}
\label{potentialunbounded}
V(\mathbf{x})=\left\{
\begin{array}{ll}
\frac{\beta}{8}\mathbf{x}^2 & \text{if } \mathbf{x}\in\Delta,\\
0 & \text{if }\mathbf{x}\notin\Delta.
\end{array}
\right.
\end{eqnarray}
Then, the Lagrange parameter $\beta$ must be strictly positive in order to obtain integrable solutions.

The set of densities with a given variance $\langle\mathbf{x}^2\rangle$ is a convex set, and since the Fisher information is a convex functional, its minimum, if it exists, is unique.
However, it is not possible to find a minimal value of the Fisher information regardless of the value of $\langle \mathbf{x}^2\rangle$: the reason is that, due to the nature of the harmonic
%removed  oscillator
potential, 
%modified here
the factor $\beta^\frac14$ acts only as a scale factor for the resulting probability density function. Then, since the Fisher information scales as $\beta^{-1/2}$, it is possible to reach arbitrarily low values of the Fisher information by modifying this scale factor. Nevertheless, since the product $F\langle\mathbf{x}^2\rangle$ is scale invariant, it is independent of $\beta$, and a minimum 
%modified could
should exist. This minimum is characterized by the following theorem.

\begin{theorem}
In the case of an unbounded domain $\Delta$, the minimum value of the Cram\'{e}r-Rao product $F\langle\mathbf{x}^2\rangle$ verifies 
\[
E^2=\frac{\beta}{16}F\langle\mathbf{x}^2\rangle
\] where $E$ is the energy of the  ground state of the quantum system with potential defined by (\ref{potentialunbounded}). It is reached by the probability density associated to the ground state of this system.
\end{theorem}

\begin{IEEEproof}
We apply the virial theorem, that ensures the existence of this minimum and points to it: for the  potential (\ref{potentialunbounded}), the virial theorem establishes that the kinetic energy $\langle T\rangle=\langle V\rangle$. Then, the total energy $E=\langle T\rangle+\langle V\rangle=2\langle T\rangle=2\langle V\rangle$, and $E^2=4\langle T\rangle\langle V\rangle$. As $\langle T\rangle=F/8$ for real wave functions, and since $\langle V\rangle=\frac{\beta}{8}\langle \mathbf{x}^2\rangle$, we have that
\begin{equation}
%\label{FVunbounded}
E^2=\frac{\beta}{16}F\langle\mathbf{x}^2\rangle.
\end{equation}
Then, the minimum value of the product $F\langle\mathbf{x}^2\rangle$ is %modified here
reached for the minimum value of the energy $E$, i.e. by the ground state of this quantum system.
\end{IEEEproof}

\subsection{Example 1}

Let us consider the case of the positive half plane
\[
\Delta=\{(x,y)\in\mathbb{R}^2, x>0\}.
\]

The Schr\"odinger equation reads
\begin{multline*}
-\frac12 \left(\frac{\partial^2 u(x,y)}{\partial x^2}+\frac{\partial^2 u(x,y)}{\partial y^2}\right)
+\frac{\beta}{8}(x^2+y^2)u(x,y)\\
=-\frac{\alpha}{8}u(x,y),
\end{multline*}
% changed here
with the constraint $u(0,y)=0 \,\, \forall y \in \mathbb{R}$.

The solutions are of the form
\begin{eqnarray*}
u_{n_1,n_2}(x,y)=\sqrt{\frac{2^{-2n_1-n_2-1}\sqrt{\beta}}{\pi  (2n_1+1)! n_2!}} e^{-\frac{\sqrt{\beta}}{4}(x^2+y^2)} \\
\times H_{2n_1+1}\left(\frac{\beta^\frac14}{\sqrt{2}}x\right) H_{n_2}\left(\frac{\beta^\frac14}{\sqrt{2}}y\right) 
\end{eqnarray*}
 where $H_{n}(x)$ is the Hermite polynomial of degree $n$. The corresponding energy levels are
\[
E_{n_1,n_2}=-\frac{\alpha}{8}=\frac{\sqrt{\beta}}{2}(2n_1+n_2+2).
\]

The density $\rho_{n_1,n_2}$ is defined as
\[
\rho_{n_1,n_2}(x,y)=u_{n_1,n_2}^2(x,y);
\]
its variance is
\[
\langle x^2+y^2\rangle_{n_1,n_2}=\frac{2}{\sqrt{\beta}}(2n_1+n_2+2),
\]
and its Fisher information 
\[
F[\rho_{n_1,n_2}]=2\sqrt{\beta}(2n_1+n_2+2).
\]

Notice that, as $\beta$ is a scale factor, the product of these two quantities does not depend on $\beta$ and equals
\[
F[\rho_{n_1,n_2}]\langle x^2+y^2\rangle_{n_1,n_2}=4(2n_1+n_2+2)^2.
\]
As predicted by the virial theorem, its minimum value is obtained for the ground state
described by the density
\[
\rho_{0,0}(x,y)=\frac{\beta}{\pi} x^{2}  e^{-\frac{\sqrt{\beta}}{2}(x^2+y^2)}
\]
and is equal to
\[
F[\rho_{0,0}]\langle x^2+y^2\rangle_{0,0}=16,
\]

\section{Bounded $D$-dimensional domains}
\label{sec3}
\subsection{Statement of the problem}
In the case of a bounded domain $\Delta$, the sign of the Lagrange multiplier $\beta$ in the equation
\begin{equation}
-\frac12\boldsymbol\nabla^2u(\mathbf{x})+\frac{\beta}{8}\mathbf{x}^2u(\mathbf{x})=-\frac{\alpha}{8}u(\mathbf{x})
\label{eq.schrodinger}
\end{equation}
cannot be fixed as in the case of unbounded domains.

As the values of the $D$-dimensional variable $\mathbf{x}$ are bounded in $\Delta$, its variance  $\langle \mathbf{x}^2\rangle$ is also bounded. Then, there must 
% modified be
exist a value of the constraint $\langle \mathbf{x}^2\rangle_{*}$ for which the minimal Fisher information $F_*$ is achieved.  This value $F_*$ would be the minimal value of the Fisher information among all the densities defined in $\Delta$.

We need the two following propositions to find this minimal value and the density that achieves it.

\begin{proposition}
Let 
\[g_\epsilon(\mathbf{x})=\epsilon u^2(\mathbf{x})+(1-\epsilon) v^2(\mathbf{x})\] with $0\le \epsilon <1$. Then a first order expansion of the Fisher information of $g_\epsilon$ is
\[
F[g_\epsilon]=F[v^2]-\beta\epsilon (\langle\mathbf{x}^2\rangle_u-\langle\mathbf{x}^2\rangle_v)+o(\epsilon^2).
\]
\end{proposition}

\begin{IEEEproof}
The Fisher information of $g_\epsilon$ is
\begin{eqnarray*}
&F[g_\epsilon]=\int_\Delta \frac{|\boldsymbol\nabla g_\epsilon(\mathbf{x})|^2}{g_\epsilon(\mathbf{x})} d\mathbf{x}\\
&=4\int_\Delta \frac{v^2(\mathbf{x})|\boldsymbol\nabla v(\mathbf{x})|^2}{g_\epsilon(\mathbf{x})} d\mathbf{x}\\
&+8\epsilon \int_\Delta \frac{u(\mathbf{x})v(\mathbf{x})\boldsymbol\nabla u(\mathbf{x})\cdot\boldsymbol\nabla v(\mathbf{x})-v^2(\mathbf{x})|\boldsymbol\nabla v(\mathbf{x})|^2}{g_\epsilon(\mathbf{x})} d\mathbf{x}\\
&+4\epsilon^2 \int_\Delta \frac{u^2(\mathbf{x})|\boldsymbol\nabla u(\mathbf{x})|^2+v^2(\mathbf{x})|\boldsymbol\nabla v(\mathbf{x})|^2}{g_\epsilon(\mathbf{x})} d\mathbf{x}\\
&+4\epsilon^2 \int_\Delta \frac{-2u(\mathbf{x})v(\mathbf{x})\boldsymbol\nabla u(\mathbf{x})\cdot\boldsymbol\nabla v(\mathbf{x})}{g_\epsilon(\mathbf{x})} d\mathbf{x}\\
%&=&4\int_\Delta \frac{v^2(\mathbf{x})|\mathbf{\nabla}v(\mathbf{x})|^2+2\epsilon v(\mathbf{x}) (u(\mathbf{x})\mathbf{\nabla}v(\mathbf{x})\cdot\mathbf{\nabla}u(\mathbf{x})-v(\mathbf{x})|\mathbf{\nabla}v(\mathbf{x})|^2)}{\epsilon u^2(\mathbf{x})+(1-\epsilon) v^2(\mathbf{x})} d\mathbf{x}\\
%&& +  4\epsilon^2 \int_\Delta \frac{u^2(\mathbf{x})|\mathbf{\nabla}u(\mathbf{x})|^2+v^2(\mathbf{x})|\mathbf{\nabla}v(\mathbf{x})|^2-2u(\mathbf{x})v(\mathbf{x})\mathbf{\nabla}u(\mathbf{x})\cdot\mathbf{\nabla}v(\mathbf{x})}{\epsilon u^2(\mathbf{x})+(1-\epsilon) v^2(\mathbf{x})} d\mathbf{x}
\end{eqnarray*}
Denoting the integrand in the two first  integrals by $G(u,v)$, we perform a Taylor expansion in terms of $\epsilon$ around $\epsilon=0$, yielding the expression
\[
G(u,v)=G_0(u,v)+\epsilon G_1(u,v)+o(\epsilon^2),
\]
where
\[
G_0(u,v)=|\boldsymbol\nabla v(\mathbf{x})|^2,
\]
and
\[
G_1(u,v)=\boldsymbol\nabla v(\mathbf{x})\cdot\boldsymbol\nabla \left(\frac{u^2(\mathbf{x})-v^2(\mathbf{x})}{v(\mathbf{x})}\right).
\]

Then,
\[
F[g_\epsilon]=F[v^2]+4\epsilon \int_\Delta \boldsymbol\nabla v(\mathbf{x})\cdot\boldsymbol\nabla \left(\frac{u^2(\mathbf{x})-v^2(\mathbf{x})}{v(\mathbf{x})}\right) d\mathbf{x}+o(\epsilon^2).
\]
An integration by parts yields:
\begin{eqnarray*}
\int_\Delta\boldsymbol\nabla v(\mathbf{x})\cdot\boldsymbol\nabla \left(\frac{u^2(\mathbf{x})-v^2(\mathbf{x})}{v(\mathbf{x})}\right) d\mathbf{x}\\
=\int_{\mathcal{D}(\Delta)} \frac{u^2(\mathbf{x})-v^2(\mathbf{x})}{v(\mathbf{x})} \boldsymbol\nabla v(\mathbf{x})\cdot \mathbf{\nu} d\gamma_\Delta\\
-\int_\Delta \frac{u^2(\mathbf{x})-v^2(\mathbf{x})}{v(\mathbf{x})} \boldsymbol\nabla^2v(\mathbf{x})d\mathbf{x}
\end{eqnarray*}
where $d\gamma_\Delta$ is the infinitesimal element of the border $\mathcal{D}(\Delta)$ of $\Delta$ and $\mathbf{\nu}$ is a vector perpendicular to it.

As $u(\mathbf{x})=0$ and $v(\mathbf{x})=0$ for $x\in \mathcal{D}(\Delta)$, this result yields
\[
F[g_\epsilon]=F[v^2]-4\epsilon\int_\Delta \frac{u^2(\mathbf{x})-v^2(\mathbf{x})}{v(\mathbf{x})} \boldsymbol\nabla^2v(\mathbf{x})d\mathbf{x}.
\]
Now we use the Schr\"odinger equation
\[
-\frac12\boldsymbol\nabla^2v(\mathbf{x})+\frac{\beta}{8}\mathbf{x}^2 v(\mathbf{x})=-\frac{\alpha}{8}v(\mathbf{x}),
\]
so that we obtain the result
\[
F[g_\epsilon]=F[v^2]-\beta\epsilon(\langle \mathbf{x}^2\rangle_u-\langle\mathbf{x}^2\rangle_v)+o(\epsilon^2).
\]
\end{IEEEproof}

\begin{proposition} 
Let us denote by $\langle \mathbf{x}^2\rangle_*$ the value of the variance that, after solving the variational problem, corresponds to the minimal value of the Fisher information $F_*$. Then   the Lagrange multiplier $\beta$ of the problem has the same sign as $\langle \mathbf{x}^2\rangle_{*} - \langle \mathbf{x}^2\rangle$, that is 
\begin{itemize}
\item if $\langle \mathbf{x}^2\rangle<\langle \mathbf{x}^2\rangle_*$ then $\beta>0$,
\item if $\langle \mathbf{x}^2\rangle>\langle \mathbf{x}^2\rangle_*$ then $\beta<0$,
\item if $\langle \mathbf{x}^2\rangle=\langle \mathbf{x}^2\rangle_*$ then $\beta=0$.
\end{itemize}
\end{proposition}

\begin{IEEEproof} 
Let $u^2(\mathbf{x})$ and $v^2(\mathbf{x})$ be two distributions with respective Fisher informations $F[u^2]$ and $F[v^2]$. According to the previous proposition, if $g_\epsilon(\mathbf{x})=\epsilon u^2(\mathbf{x})+(1-\epsilon) v^2(\mathbf{x})$. Then,
\begin{equation}
F[g_\epsilon]=F[v^2]-\beta\epsilon (\langle\mathbf{x}^2\rangle_u-\langle\mathbf{x}^2\rangle_v)+o(\epsilon^2).
\label{eq.fgepsilon}
\end{equation}
The convexity of the Fisher information gives
\[
F[g_\epsilon]<\epsilon F[u^2]+(1-\epsilon)F[v^2]
\]
Let us now take $\langle x^2\rangle_u=\langle x^2\rangle_*$. Then, $F[u^2]=F_*$ is the minimum Fisher information, so we have the majoration
\[
\epsilon F[u^2]+(1-\epsilon)F[v^2]\le F[v^2].
\]
Therefore, $F[g_\epsilon]<F[v^2]$, and, as a consequence, from (\ref{eq.fgepsilon}) and the previous inequality, we obtain
\[
\beta\epsilon(\langle\mathbf{x}^2\rangle_u-\langle\mathbf{x}^2\rangle_v)>0,
\]
which gives the first and second cases in the proposition. The third case follows by continuity.

\end{IEEEproof}

For $\beta=0$, the Schr\"odinger equation (\ref{eq.schrodinger}) becomes
\begin{equation}
\label{schroedinger}
-\frac12\boldsymbol\nabla^2u(\mathbf{x})=-\frac{\alpha}{8}u(\mathbf{x})
\end{equation}
for $\mathbf{x}\in\Delta$, with $u(\mathbf{x})=0$ $\forall \mathbf{x}\in \mathcal{D}(\Delta)$,
that is the equation of the $D$-dimensional infinite potential well defined in $\Delta$.

We can now characterize the minimum Fisher information in the case of a bounded domain.
\begin{theorem}
The minimum Fisher information in a bounded domain of $\mathbb{R}^D$ is reached by the probability density associated to the ground state of the quantum system whose potential is the infinite well in this domain.
\end{theorem}
\begin{IEEEproof}
Notice that if $u(\mathbf{x})$ is a solution of equation (\ref{schroedinger}), then $|u(\mathbf{x})|$ is also a solution. Since the solution of the problem of finding the minimum value of $F$ for a given value of $\langle \mathbf{x}^2\rangle$ must be unique, we conclude that $u(\mathbf{x})=|u(\mathbf{x})|$, so $u(\mathbf{x})>0$ for $\mathbf{x}\in\Delta$. But the only eigenstate of the previous Schr\"odinger equation that does not have any zero is the ground state. Thus, the density that minimizes the Fisher information in a bounded domain is that associated to the ground state of the corresponding infinite well defined on that domain.
\end{IEEEproof}

\subsection{Example 2}

Let us consider the case of the rectangular domain
\[
\Delta=\{(x,y)\in\mathbb{R}^2,0<x<2,-1<y<2\}.
\]

The Schr\"odinger equation reads
\[
-\frac12 \left(\frac{\partial^2 u(x,y)}{\partial x^2}+\frac{\partial^2 u(x,y)}{\partial y^2}\right)=-\frac{\alpha}{8}u(x,y),
\]
with the constraint $u(x,y)=0$ if $(x,y)\notin\Delta$.

The solutions are of the form
% added indices n_1 and n_2 here
\[
u_{n_{1},n_{2}}(x,y)=\sqrt{\frac32}\sin\left(\frac{n_1\pi}{2}x\right)\sin\left(\frac{n_2\pi}{3}(y+1)\right)
\]
with energy levels
\[
E_{n_1,n_2}=\frac{\pi^2}{8}\left(n_1^2+\frac49 n_2^2\right).
\]

The associated density is
\[
\rho_{n_1,n_2}(x,y)=u_{n_{1},n_{2}}^{2}(x,y).
\]
The Fisher information of this density is
\[
F[\rho_{n_1,n_2}]=\pi ^2\left( n_1^2+\frac49 n_2^2\right).
\]
(Notice that $E_{n_1,n_2}=F[\rho_{n_1,n_2}]/8$ as $\langle V\rangle=0$).

The ground state, corresponding to $n_1=n_2=1,$ reads
\[
\rho_{1,1}(x,y)=\frac32 \sin^2\left(\frac{\pi}{2}x\right)\sin^2\left(\frac{\pi}{3}(y+1)\right),
\]
and is the only eigenstate without zeros.
Its Fisher information
\[
F[\rho_{1,1}]=\frac{13\pi^2}{9}
\]
is the minimum Fisher information that can be achieved for any density defined in the domain $\Delta$.

The expectation value of $\langle \mathbf{x}^2\rangle$ has the value
\[
\langle x^2+y^2\rangle_{n_1,n_2} = \frac73-\frac{1}{2\pi^2}\left(\frac{4}{n_1^2}+\frac{9}{n_2^2}\right).
\]
Then,
\[
\langle x^2+y^2\rangle_{1,1}F[\rho_{1,1}]=\left( \frac73-\frac{13}{2\pi^2}\right)\frac{13\pi^2}{9} \approx 23.875
\]
is found to be the minimal value of the Cram\'er-Rao product.

\section{Conclusions}
In summary, we have studied the probability densities
yielding the minimum value of the Cramer-Rao product 
in general $D$-dimensional unbounded domains and 
the densities minimizing the Fisher information
itself in  bounded domains $\Delta\in {\cal R}^D$. 
The present developments constitute a substantial 
generalization of previous results obtained by Becher 
and Vignat for the one dimensional case \cite{bercher.is09}.  In the 
case of unbounded domains the density optimising the 
Cramer-Rao product turns out to be the one corresponding 
to the ground state of the ($D$-dimensional) harmonic 
oscillator in the domain. On the other hand, the optimal
density minimising the Fisher information in a bounded 
domain $\Delta$  is given by the ground state
of a quantum free particle confined within a rigid box
described by the boundary of $\Delta$. This may have 
applications for the study of important quantum systems 
such as quantum billiards. 
 
It is intriguing that the above mentioned optima are achieved
by the ground states of two systems that are among the
most basic (and most important) in quantum physics. This 
constitutes new evidence pointing towards the fundamental
role played by Fisher information both in  quantum mechanics
and information theory, and particularly in the interface 
between these two fields.

\section*{Acknowledgements}

This work has been partially supported by the MICINN grant FIS2008-2380 and the grants FQM-2245 and FQM-4643 of the Junta de Andaluc\'{\i}a. JSD, ARP and PSM belong to the research group FQM-207. C. V. thanks Pr. Dehesa for his kind invitation to Granada in November 2009 that allowed to initiate this work.

\bibliographystyle{IEEEtran}
\bibliography{cramer}

\end{document}